\documentclass{article}
\usepackage{amssymb}

\input{tcilatex}

\begin{document}

\title{Betweenness of partial orders}
\author{Bruno Courcelle \\
LaBRI, CNRS and Bordeaux University, \\
33405\ Talence, France\\
courcell@labri.fr}
\maketitle

\begin{quote}
\textbf{Abstract: }We construct a monadic second-order sentence that
characterizes the ternary relations that are the betweenness relations of
finite or infinite partial orders.\ We prove that no first-order sentence
can do that.\ We characterize the partial orders that can be reconstructed
from their betweenness relations. We propose a polynomial time algorithm
that tests if a finite relation is the betweenness of a partial order.

\bigskip 
\end{quote}


\bigskip 

{\LARGE Introduction}

\bigskip 

\emph{Betweenness} is a standard notion in the study of structures such as
trees, partial orders and graphs. It is defined as the ternary relation $%
B(x,y,z)$ expressing that an element $y$ is \emph{between} $x$ and $z$, in a
sense that depends on the considered structure.\ This relation is easy to
understand and axiomatize in first-order (FO) logic for \emph{linear orders}%
.\ In particular, a linear order can be uniquely described, up to reversal,
from its betweenness relation.\ However, the notion of \emph{partial
betweenness\footnote{%
The question is whether a given ternary relation $B$ is included in the
betweenness relation of a linear order.\ The corresponding problem is
NP-complete. This question is not related with the betweenness of partial
orders.}} raises some difficult algorithmic and logical problems (\cite%
{CouEng}, Chapter 9).

Betweenness in \emph{partial orders} is axiomatized in \cite{Lih} by an
infinite set of FO sentences that cannot be replaced by a finite one, as we
will prove. In this article, we axiomatize betweenness in partial orders by
a \emph{single monadic second-order} (MSO) sentence.\ We characterize the
partial orders that are uniquely reconstructible, up to reversal, from their
betweenness relations. We show that an MSO\ formula can describe \emph{some}
partial order $P$\ such that $B_{P}=B_{Q}$ from the betweenness relation $%
B_{Q}$ of a partial order $Q$: this definition yields a partial order that
may be a proper suborder of $Q$. We give a polynomial time algorithm to test
if a finite ternary structure is the bewteenness relation of some partial
order and to produce relevant partial orders if this is possible.

Several notions of betweenness in \emph{graphs} have also been investigated
and axiomatized.\ We only refer to the survey \cite{Cha+} that contains a
rich bibliography. Another reference is \cite{Chv}. In previous articles we
have studied betweenness in finite or infinite trees, and also in \emph{%
generalized trees}, defined as the partial orders such that the set of
elements larger than any one is linearly ordered \cite%
{CouRwd,CouLMCS,CouBOtrees,CouYuri}. The corresponding betweenness relations
are defined from partial orders, but not as in \cite{Lih} and in the present
article.

This work contributes to the understanding of the expressive power of
monadic second-order logic in finite and infinite graphs and related
relational structures. We refer to \cite{Cou15,CouEng} for monadic
second-order logic.

\section{Definitions and known results}

All partial orders, graphs and relational structures are finite or countably
infinite.

\textbf{Definitions 1\ : }\emph{Betweenness.}

To shorten writings $\neq (x_{1},x_{2},...,x_{n})$ means that $%
x_{1},x_{2},...,x_{n}$ are pairwise distinct.

(a) \emph{Betweenness in linear orders}

Let $L=(V,\leq )$ be a linear order.\ Its \emph{betweenness relation} $B_{L}$
is the ternary relation on $V$ defined by :

\begin{quote}
$B_{L}(x,y,z):\Longleftrightarrow x<y<z$ or $z<y<x.\ $
\end{quote}

The following properties hold for $B=B_{L}$ and all $x,y,z,u\in V$:

\begin{quote}
B1 : $B(x,y,z)\Rightarrow \neq (x,y,z).$

B2 : $B(x,y,z)\Rightarrow B(z,y,x).$

B3 : $B(x,y,z)\Rightarrow \lnot B(x,z,y).$

B4 : $B(x,y,z)\wedge B(y,z,u)\Rightarrow B(x,y,u)\wedge B(x,z,u).$

B5 : $B(x,y,z)\wedge B(x,u,y)\Rightarrow B(x,u,z)\wedge B(u,y,z).$

B6 : $\neq (x,y,z)\Rightarrow B(x,y,z)\vee B(x,z,y)\vee B(y,x,z).$
\end{quote}

We get an\ axiomatization by finitely many universal first-order sentences:
if a ternary structure $S=(V,B)$ satisfies these properties, then $B=B_{L}$\
for a linear order $L=(V,\leq )$.\ We will say that the class of betweenness
relations of linear orders is \emph{first-order} (\emph{FO}) \emph{definable}%
.\ The order $L=(V,\leq )$ and its reversal $L^{rev}:=(V,\geq )$ are the
only ones whose betweenness relation is $B_{L}$, see \cite%
{CouRwd,CouLMCS,CouEng}.\ We will say that $\leq $ is \emph{uniquely }%
defined, \emph{up to reversal} (written \emph{u.t.r}.), or \emph{%
reconstructible} from its betweenness relation.

(b) \emph{Betweenness in partial orders}

The \emph{betweenness relation} $B_{P}$ of a partial order $P=(V,\leq )$ --
we will also say a \emph{poset\footnote{%
This is an inelegant but short terminology for \emph{partial order}, \emph{%
partial ordering} or \emph{partially ordered set}.}} -- is the ternary
relation on $V$ defined, as in (a), by :

\begin{quote}
$B_{P}(x,y,z):\Longleftrightarrow x<y<z$ or $z<y<x.$
\end{quote}

We denote by $Bet(P)$ the ternary structure $(V,B_{P})$. For all $%
x,y,z,u,v\in V$, the relation $B=B_{P}$ satisfies Properties B1 to B5\
together with:

\begin{quote}
X : $B(x,y,z)\wedge B(u,y,v)\Rightarrow B(x,y,u)\vee B(x,y,v),$

F : $B(x,y,z)\wedge B(y,u,v)\Rightarrow B(x,y,u)\vee B(z,y,u),$
\end{quote}

and an infinite set $\mathcal{O}$ of properties expressed by universal
first-order sentences.\ The notation is borrowed to the article by Lihova 
\cite{Lih} who proved, conversely, that if a ternary structure $S=(V,B)$
satisfies these properties, then $B=B_{P}$\ for a poset $P=(V,\leq )$ and,
of course for its reversal $P^{rev}:=(V,\geq )$. We will prove that no
finite set of first-order sentences can characterize betweenness in posets.
Our proof will use the following examples.

(c) A \emph{B-cycle} is a ternary structure $(V,B)$ such that $%
V=\{a_{1},a_{2},...,a_{n},b_{1},b_{2},$ $...,b_{n}\}$, $n\geq 2$, and $B$\
consists of the triples $%
(a_{1},b_{1},a_{2}),(a_{2},b_{2},a_{3}),...,(a_{n-1},b_{n-1},$\ $%
a_{n}),(a_{n},b_{n},a_{1})$\ and their inverse ones, $%
(a_{2},b_{1},a_{1}),(a_{3},b_{2},a_{2}),...$ so that B2 is satisfied. This
structure satisfies Properties B1-B5.\ If $n$ is even, then $B=B_{P}$ where $%
P=(V,\leq )$ is the poset such that :

\begin{quote}
$%
a_{1}<b_{1}<a_{2}>b_{2}>a_{3}<b_{3}<a_{4}>...a_{n-1}<b_{n-1}<a_{n}>b_{n}>a_{1}, 
$
\end{quote}

and no other inequality holds except by transitivity (\emph{e.g.} $%
a_{1}<a_{2}$). If $n$ is odd, no such partial order does exist (cf.\ Lemma
14).\ Consider for example the case $n=3$.\ A partial order $P$ such that $%
a_{1}<b_{1}$ and $B_{P}=B$\ must verify $b_{1}<a_{2}>b_{2}>a_{3}<b_{3}<a_{1}$
but then, we would have $(b_{3},a_{1},b_{1})$ in $B$, which is not assumed.
The set $\mathcal{O}$ excludes these odd $B$-cycles.

We will prove that the class of betweenness relations of partial orders is
monadic second-order (MSO) definable without using the set $\mathcal{O}$. We
will also identify the partial orders that can be reconstructed \emph{u.t.r}%
. from their betweenness relations, independently of any logical
description. We refer to \cite{Cou15,CouEng} for first-order and monadic
second-order logic.

\textbf{Definitions 2 }: \emph{Ternary structures and their Gaifman graphs.}

(a) A \emph{ternary structure} is a pair $S=(V,B)$ such that $B\subseteq
V^{3}.$ Its \emph{Gaifman graph\footnote{%
This graph is defined similarily for arbitrary relational structures, not
only for ternary ones.}} is $Gf(S):=(V,E)$ where there is an edge $u-v$ in $%
E $ if and only if $u\neq v$ and $u$ and $v$ belong to a same triple in $B$.

We say that $S$\ is \emph{connected} if $Gf(S)$ is. If $Gf(S)$ is not
connected, then $S$ is the union of the pairwise disjoint induced structures 
$S[X]:=(X,B\cap X^{3})$, called the \emph{connected components} of $S$,
where the sets $X$ are the vertex sets of the connected components of $Gf(S)$%
.

(b) If $P=(V,\leq )$ is a partial order, its \emph{comparability graph} $%
Comp(P)$ having vertex set $V$\ and an edge $u-v$ if and only if $u$ and $v$
are different and \emph{comparable}, \emph{i.e.}, $u<v$ or $v<u$.\ It is the
Gaifman graph of the binary structure $P$.

We say that $P$\ is \emph{connected} if $Comp(P)$ is. If $Comp(P)$ is not
connected, then $P$ is the union of the pairwise disjoint posets $%
P[X]:=(X,\leq \cap X^{2})$ where the sets $X$ are the vertex sets of the
connected components of $Comp(P)$.

We have $Gf(Bet(P))\subseteq Comp(P)$.\ The inclusion may be proper.

If $Gf(Bet(P))$ is connected, then so is $Comp(P),$ but not necessarly
conversely, because $Gf(Bet(P))$ has no edge if $P$ has no chain $x<y<z$.

\bigskip

\textbf{Example 3} : Here is an example where $Gf(Bet(P))\subset Comp(P)$.\
Let $P=(V,\leq )$ where $V=\{a,b,c,d,e,f\}$ and $\leq $\ is generated by $%
a<b<c<d,$\ $e<c,e<f,b<f,$ reflexivity and transitivity.\ The edge $e-f$ of $%
Comp(P)$ is not in $Gf(Bet(P))$ because $e$ and $f$ do not belong to any
chain of size 3. If we remove the clause $e<f$, the resulting partial order
has the same betweenness relation as $P$ and $Gf(Bet(P))=Comp(P)$.\ We will
generalize this observation in Proposition 6.

\section{Betweenness in partial orders}

\bigskip

\textbf{Definitions 4 : }\emph{Chains, antichains and B-minimality}.

Let $P=(V,\leq )$ be a partial order.\ 

(a) A \emph{chain} (resp. an \emph{antichain}) is a subset $X$\ of $V$\ that
is linearly ordered (resp. where any two elements are incomparable.)\ Its 
\emph{size} is $\left\vert X\right\vert \in \mathbb{N}\cup \{\omega \}.$
Maximality of chains and antichains is understood for set inclusion.

We use $A_{P}(x,y,z)$ to abreviate $B_{P}(x,y,z)\vee B_{P}(x,z,y)\vee
B_{P}(y,x,z),$ meaning that $x,y$ and $z$ belong to a chain of size at least
3.

(b) We say that $P$\ is \emph{B-minimal} if $Gf(Bet(P))=Comp(P)$,
equivalently, if every two comparable elements belong to a chain of size at
least 3, or, as we will see, that it is the unique minimal poset\ $P$\ such
that $Gf(Bet(P))=Gf(Bet(Q))$ for some poset $Q$, where posets are related by
inclusion of the defining binary relations.

(c) We define $Min(P)$ and $Max(P)$ as the sets of minimal and maximal
elements respectively of a poset $P$.\ They are its \emph{extremal elements}%
.\ An element is \emph{isolated} if it is so in $Comp(P)$, equivalently, if
it belongs to $Min(P)\cap Max(P)$.

(d) In a ternary structure $S=(V,B)$ that satisfies Properties B1,B2 and B3
(in order to avoid uninteresting cases), we say that an element $x$ is \emph{%
extremal} if $B(y,x,z)$ does not hold for any $y,z.$ The extremal elements
of a structure $Bet(P)$ are the extremal elements of $P$.$\square $

\textbf{Example 5:} \ Let $P=(V,\leq )$ where $V=\{a,b,c,d,e,f\}$ and $\leq $%
\ is generated by $a<b<c,$\ $d<e<f,d<c,$ reflexivity and transitivity. Then $%
Comp(P)$ is connected but $Gf(Bet(P))$ is not.\ The only orderings on $V$
that yield $Comp(P)$ as comparability graph are $P$ and $P^{rev}$.\ The
graph $Gf(Bet(P))$\ has two connected components with vertex sets $\{a,b,c\}$%
\ and $\{d,e,f\}$. From it, one obtains 4 orderings on $V$ that yield the
betweenness structure $Bet(P).$ We will develop this observation. $\square $

\textbf{Proposition 6} : Let $P=(V,\leq )$ be a poset and $\leq ^{\prime }$
be defined by :

\begin{quote}
$x\leq ^{\prime }y$ if and only if $x=y$ or

$x<y$ and, if $x\in Min(P)$ and $y\in Max(P)$, then $x<z<y$ for some $z$.\ 
\end{quote}

The poset $\widetilde{P}:=(V,\leq ^{\prime })\subseteq P$ is B-minimal and $%
Bet(\widetilde{P})=Bet(P)$. It is the unique minimal poset\ $Q$ such that $%
Bet(Q)=Bet(P)$ and $Q\subseteq P$.

\textbf{Proof} : We have $\leq ^{\prime }$ $\subseteq \ \leq $ , hence, $%
\widetilde{P}\subseteq P$.\ If $x<^{\prime }y<^{\prime }z$, then $x<y<z$,
hence $x<z$, and if $x\in Min(P)$ and $z\in Max(P)$, we have $y$ between
them, hence $x<^{\prime }z$.\ \ Reflexivity and antisymmetry are clear and
so we have a partial order.

We have $Bet(\widetilde{P})\subseteq Bet(P)$. However, if $x<y<z,$ we have $%
x<^{\prime }y<^{\prime }z$ by the definitions. Hence, $\widetilde{P}$ and $P$
have the same chains of size at least 3.\ In particular, $Bet(\widetilde{P}%
)=Bet(P)$.

If $\widetilde{P}$ is not B-minimal, there are $x,y$ such that $x<^{\prime
}y $ and $x$ and $y$ do not belong to any chain of size 3 in $\widetilde{P}$%
, whence, in $P$.\ As $x<y$\ this implies that $x\in Min(P)$ and $y\in
Max(P) $, but we have $x<z<y$ for some $z,$ hence we have $x<^{\prime
}z<^{\prime }y $, which contradicts the assumption that $x$ and $y$ do not
belong to a chain of size 3 in $\widetilde{P}$.

Assume that\ $Q=(V,\leq _{Q})\subseteq P$ and $Bet(Q)=Bet(P)$. If $%
x<^{\prime }y$, then, we have $x<_{Q}y$: to prove this, we observe that the
defintions yield $x<y<z$ or $z<x<y$ or $x<z<y$ of some $z$.\ In the first
case, $(x,y,z)\in B_{P}=B_{Q}$ \ , hence $x<_{Q}y<_{Q}z$ because $Q\subseteq
P$, and so $x<_{Q}y.$ The proofs are similar for the two other cases.\ Hence 
$\widetilde{P}\subseteq Q$.\ Hence, $\widetilde{P}$ is the unique minimal
poset\ $Q$ such that $Bet(Q)=Bet(P)$ and $Q\subseteq P$. $\square $

\bigskip

A partial order $P$ is \emph{B-reconstructible} (that is \emph{%
reconstructible from its betweenness relation}) if $P$ and $P^{rev}$ are the
only ones whose betweenness structure is $Bet(P)$.

\textbf{Theorem 7 }: A partial order is B-reconstructible if and only if it
is B-minimal and, either it is connected or it has exactly two connected
components that are one without extremal elements and an isolated element.\
A finite B-reconstructible partial order is B-minimal and connected.

\bigskip

\textbf{Proof }: We first recall the case of a linear order $L=(V,\leq )$
from\ \cite{CouLMCS} that is connected and B-minimal. If $a<b$ in $V$, then $%
\leq $ is the unique linear order\footnote{%
This order is FO definable in the structure $(V,B_{L},a,b)$.}\ such that $%
a<b $ and whose betweenness relation is $B_{L}$.\ 

"If" Let $P=(V,\leq )$\ be connected and B-minimal, and let $B$ be its
betweenness relation.\ Let $C$\ be a maximal chain (it has size at least 3)
and $a,b$\ in $C$\ such that $a<b$. The linear order $(C,\leq )$\ is
uniquely determined by $B\cap C^{3}$\ and the pair $(a,b)$.

Consider another maximal chain $D$. If $\left\vert C\cap D\right\vert \geq 2$%
, we say that $C$ and $D$ \emph{merge}.\ Then, the order on $D$\ is uniquely
determined from that on $C$ and the relation $B\cap (C\cup D)^{3}$.\ 

If $C\cap D=\{c\}$, then either $Min(C)=Min(D)=c$ or $Max(C)=Max(D)=c$, and
we say that $C$ and $D$ \emph{join}.\ The order $\leq $ on $D$\ is uniquely
determined from that on $C$ and $B\cap (C\cup D)^{3}$ because we have, for
any $d\in D$ :

\begin{quote}
$c<d$ if $c<a$ and $\lnot B(a,c,d)$ and, dually,

$d<c$ if $a<c$ and $\lnot B(a,c,d).$
\end{quote}

Then, the ordering on $D$ between $c$ and $d$ determines in a unique way the
ordering on $D$. Hence, if there is a sequence of maximal chains $%
C=C_{1},C_{2},...,C_{n}$ such that $C_{i}$ and $C_{i+1}$\ merge or join for
each $i$, then the ordering on all of them is determined from the ordering
on $C$.\ 

Finally, we prove that any two comparable elements of $V$ are related in a
unique way, provided the ordering of a maximal chain $C$\ is fixed. Let $%
a\in C$ and $a=b_{0}-b_{1}-b_{2}-...-b_{n}$ be a path in the comparability
graph. There is a sequence of maximal chains $C=C_{1},C_{2},...,C_{n}$ such
that, for each $i,$ $C_{i}$ and $C_{i+1}$\ merge or join, and $b_{i}$ and $%
b_{i+1}$ are in $C_{i+1}$. The proof is by induction on $n$. We only
consider the first step.

Let $C$\ be a maximal chain containing $a$, and $b=b_{1}\notin C$ be such
that $a<b$. There is a maximal chain $D$\ that contains $a$ and $b$.\ If $C$
and $D$\ merge, we are done.\ Otherwise, $C\cap D=\{a\}$. If there is $c\in
C $ such that $c<a$, then we can replace $D$ by a maximal chain $D^{\prime }$
containing $c,a$ and $b$, and it merges with $C$; otherwise, $a=Min(C)$ and
by the maximality of $D$, we have $a=Min(D)$, hence $C$ and $D$ join. The
proof is similar if $b<a$. We let $C_{1}$ be $D$ or $D^{\prime }$ in the
second case.

It follows that the partial order on $V$ is uniquely determined from $B$ and
the linear order on $C$, because if $x,y$ are adjacent in $Comp(P)$, then
there is a sequence $C=C_{1},C_{2},...,C_{n}$ such that $C_{i}$ and $C_{i+1}$%
\ merge or join, and $x,y\in C_{n}.\ $The relation $x<y$ or $y<x,$ is thus
determined in a unique way.

As there are on $C$\ exactly two linear orders compatible with $B$, there
are exactly two partial orders on $V$, $P$ and $P^{rev}$ whose betweenness
relation is the given $B_{P}$.

If $P$\ is B-minimal and has one connected component $X$\ without extremal
elements and one isolated element $u$, then, the set $X$\ can be ordered in
exactly two ways, and $u$ must be isolated in any poset $Q$ such that $%
Bet(Q)=Bet(P)$, otherwise, any ordering $u<v$ or $v<u$ for $v\in X$ would
add triples to $B_{P}$.

"Only if" Let $P=(V,\leq )$\ be not B-minimal.\ There are $x,y\in V$ that do
not belong to any chain of size 3, and such that $x<y$.\ Hence, $x\in Min(P)$
and $y\in Max(P).$ If we remove from $\leq $ the pair $(x,y)$, as in the
definition of $\widetilde{P}$, we obtain a poset with same betweenness
relation as $P$ and that is not the reversal of $P$.

If $P$\ is B-minimal and has two connected components that are not
singletons, then each of them can be ordered in two ways while giving the
same betweenness relation as $P$.\ Hence, $V$ can be ordered in at least 4
ways, hence, $P$ is not B-reconstructible.

Let $P$\ be B-minimal.\ If it has two isolated elements $u$ and $v$ we can
order them $u<v,$ or $v<u$ or leave them incomparable, without modifying $%
Bet(P)$.\ If $P$ has one isolated element $u$ and a component having an
extremal element $w$, then we can order $V$ so that $w$ is maximal and then,
we define $u<w$ or leave $u$ isolated. Hence, $P$ is not B-reconstructible.

As a connected component without extremal elements must be infinite, the
last assertion holds. $\square $

Next we prove definability results in MSO\ logic. We let $\mathcal{B}_{PO}$
be the class of structures $Bet(P)$ for posets $P$.

\textbf{Theorem 8}: The class $\mathcal{B}_{PO}$ is MSO\ definable. There is
a pair of monadic second-order formulas that defines, for each structure $S$
in $\mathcal{B}_{PO}$,\ some partial order $P$ such that $S=Bet(P)$.

\textbf{Definition 9\ : }\emph{Cut of a partial order.}

A \emph{cut} of a poset $P=(V,\leq )$\ is a partition $(L,U)$ of $V$ such
that :

\begin{quote}
(i) $L$ is downwards closed and $U$ is upwards closed,

(ii) Every maximal chain meets $L$ and $U$.
\end{quote}

Note that $(U,L)$ is a cut of the reversal $(V,\geq )$ of $P$.\ These cuts
are Dedekind cuts in linear orders.

\textbf{Lemma 10 :} Every poset $P$ without isolated element has a cut.

\textbf{Proof: \ }We let $A$\ be a maximal antichain of $P$. There exists
one that is constructible from an enumeration\footnote{%
We consider finite or countably infinite sets that are \emph{effectively
given}, see \cite{CouEng}, hence, that have some explicit or implicit
enumeration. The Choice Axiom can also be used to assert the existence of an
enumeration.} $v_{1},v_{2},..$. of $V.$ We define

\begin{quote}
$U:=\{x\in V-Min(P)\mid y\leq x$ for some $y\in A\}$ and $L:=V-U$.
\end{quote}

We prove that $(L,U)$ is a cut. From the definition, $U$ is upwards closed,
and so, $L$ is downwards closed. Let $C$ be a maximal chain: it contains a
unique element $a\in A$.\ If $a=Min(C)$, then $a\in Min(P)\subseteq L$.\ As $%
a$ is not isolated in $P$, $a\neq Max(C)$, hence, we have $a<x$ for some $%
x\in C$, and so $x\in U$. Otherwise, we have $y<a$ for some $y\in C$, hence $%
a\notin Min(P)$ and so $a\in U$ and $y\notin U$.\ Hence, $(L,U)$ is a cut. $%
\square $

\textbf{Proof of Theorem 8\ }:

It follows from Proposition 6 that $\mathcal{B}_{PO}$ is the class of
structures $Bet(P)$ for B-minimal posets $P$.\ 

\emph{First part}: We first characterize the structures $Bet(P)$ for
B-minimal posets $P$ without isolated elements.\ 

Let $P=(V,\leq )$ be B-minimal without isolated elements, and let $(L,U)$ be
a cut of it.

We claim that $\leq $ can be defined from $L,U$ and its betweenness relation 
$B_{P}$ by FO\ formulas\footnote{%
We allow free set variables, here $L$, in FO\ formulas. We make explicit the
dependence on the relation $B$.}.

\emph{Claim 1} : For $x,y\in V,$ we have $x<y$ if and only if one of
following conditions holds, for $B:=B_{P}$:

(i) $x\in L$, $y\in U$, and $B(x,y,z)\vee B(x,z,y)\vee B(z,x,y)$ holds for
some $z$,

(ii) $x,y\in L$ and $B(x,y,w)$ holds for some $w\in U$,

(iii) $x,y\in U$ and $B(w,x,y)$ holds for some $w\in L$.

\emph{Proof} : Let $x,y$ be such that $x<y$.\ As $P$ is B-minimal, $x$ and $%
y $ belong to a maximal chain $C$ of size at least 3.\ This chain contains
some $z$ such that $x<y<z$, $x<z<y$ or $z<x<y$ and meets $L$ and $U$.\ 

If $x\in L$ and $y\in U$, then each of these three cases can hold and yields
respectively $B(x,y,z)$, $B(x,z,y)$ or $B(z,x,y).$If $x,y\in L$ , then,
since $C$ meets $U$, we have some $w\in U$ such that $x<y<w$ and $B(x,y,w)$
holds. If $x,y\in U$ then, since $C$ meets $L$, we have some $w\in L$ such
that $w<x<y$ and $B(w,x,y)$ holds. Hence, we have one of the exclusive cases
(i), (ii) or (iii).$\square $

Let $\varphi (B,L,x,y)$ be the FO\ formula expressing Conditions (i),(ii)
and (iii) of the claim in a ternary structure $S=(V,B)$, where $L\subseteq V$%
, and $U$ is defined as $V-L$. Then, there exists an FO formula $\psi (B,L)$
expressing the following:

(a) $B$ satisfies properties B1, B2 and B3, and every element of $V$\
belongs to some triple in $B$.

(b) $L$\ and its complement $U:=V-L$ are not empty,

(c) the binary relation $\{(x,y)\in V\times V\mid S\models \varphi
(B,L,x,y)\}$ is a strict partial order $<$ that is B-minimal\footnote{\emph{%
I.e.}, whose associated partial order $\leq $\ is B-minimal.},

(d) $B$\ is the betweenness relation of $<.$

Finally, we let $\theta (B)$ be the MSO\ sentence $\exists L.\psi (B,L)$.

\emph{Claim 2} : For a ternary structure $S=(V,B)$, we have $S\models \theta
(B)$ if and only if $S=Bet(P)$ for a B-minimal poset $P=(V,\leq )$ and $S$\
has no isolated elements.

\emph{Proof:} "If" Let $P=(V,\leq )$ be B-minimal, $B=B_{P}$ be its
betweenness relation, and assume that $Bet(P)$ has no isolated elements.\
The poset $P$ has none either and has a cut $(L,U)$. Properties (a) and (b)
hold by the definitions.\ By Claim 1, $\varphi (B,L,x,y)$ defines the strict
partial order $<$ and so, (c) and (d) hold.\ So we have $S\models \theta (B)$%
.

Conversely, assume that $B$\ and $L$ satisfy $\psi (B,L)$. Properties (c)
and (d) hold hence $B=B_{P}$\ for the strict and B-minimal partial order $P$
defined by $\varphi (B,L,x,y).$ By Property (a), $S$\ has no isolated
elements. $\square $

Hence, to define a B-minimal partial order $P$ such that $Bet(P)=S$ where $S$
satisfies $\theta (B)$, we use the following MSO\ formulas:

\begin{quote}
$\psi (B,L)$ intended to select in $S=(V,B)$ an appropriate set $L$.\ 

$\varphi (B,L,x,y)$ that defines the partial order in terms of $L$ assumed
to satisfy $\psi (B,L)$.
\end{quote}

If $P$ is a partial order such that $Bet(P)$ has no isolated elements, then, 
$Bet(P)\models \theta (B)$ because, by Proposition 6, $Bet(P)=Bet(\widetilde{%
P})$\ where $\widetilde{P}$\ is B-minimal.\ But the formula $\varphi
(B,L,x,y)$ defines in the structure $Bet(P)$ partial orders \ $Q$ such that $%
Bet(Q)=Bet(P)$.\ 

\emph{Second part }: \ Let be given a structure $S=(V,B)$.\ It is the union
of its connected components $S[X]$ where the sets $X$\ are the vertex sets
of the connected components of $Gf(S)$.\ There is nothing to verify for the
components which are isolated elements.\ The others can identified by an
MSO\ formula $\gamma (X),$ cf.\ \cite{CouEng}.\ Then, $S\in \mathcal{B}_{PO}$
if and only if each of these components satisfies $\theta (B)$.\ For this
purpose, we translate $\theta (B)$ into a formula $\theta ^{\prime }$ such
that, for every subset $X$ of $V$,

\begin{quote}
$S\models \theta ^{\prime }(B,X)$ if and only if $S[X]\models \theta (B\cap
X^{3})$.
\end{quote}

This is a classical construction called \emph{relativization of
quantifications} to a set $X$, see $e.g$. \cite{CouEng}. Hence, a structure $%
S$\ belongs to $\mathcal{B}_{PO}$ if and only if:

\begin{quote}
$S\models \forall X.(\gamma (X)\Longrightarrow \theta ^{\prime }(B,X))$.\ 
\end{quote}

Similarily, $\psi (B,L)$ can be transformed into $\psi ^{\prime }$ such
that, if $X,L\subseteq V$, then:

\begin{quote}
$S\models \psi ^{\prime }(B,L\cap X,X)$ if and only if $S[X]\models \psi
(B\cap X^{3},L\cap X).$
\end{quote}

It follows that, from a set $L\subseteq V$\ such that:

\begin{quote}
$S\models \forall X.(\gamma (X)\Longrightarrow \psi ^{\prime }(B,L\cap X,X))$%
,
\end{quote}

one can define partial orders for the components $S[X]$.\ We transform $%
\varphi (B,L,$ \ $x,y)$\ into $\varphi ^{\prime }(B,L,X,x,y)$ that defines a
partial order on $X$, by using $L\cap X\ $. Then, from $L$ as above, one
obtains a strict partial order on $V$\ defined by :

\begin{quote}
$x<y:\Longleftrightarrow S\models \exists X.(x,y\in X\wedge \gamma (X)\wedge
\varphi ^{\prime }(B,L,X,x,y))$
\end{quote}

whose betweenness relation is $B$. \ This completes the proof of the
theorem. $\square $

\textbf{Remark 11}: A B-minimal poset $P=(V,\leq )$ has several cuts $(L,U)$%
. However, from the structure $Bet(P)$, they yield only two orders whose
betweenness relation is $Bet(P)$. If $Gf(Bet(P))$ has $n$ connected
components that are not singleton, the formulas\ $\psi ^{\prime }$ and $%
\varphi ^{\prime }$\ define the $2^{n}$ partial orders whose betweenness is $%
Bet(P)$.

\textbf{Remark 12} : A partial order is reconstructible, \emph{u.t.r}., from
its comparability graph if and only if this graph is \emph{prime}.\ This
notion is relative to the theory of modular decomposition.\ Furthermore,
there is an MSO\ formula that defines the two transitive orientations of a
prime comparability graph $G$, equivalently, the two partial orders $P$\
such that $G=Comp(P)$.\ More generally, the class of comparability graphs is
MSO definable, and primality is MSO\ definable.\ These results are proved in
Section 5 of \cite{Cou15}. They concern finite and countably infinite
partial orders.

\section{Finite partial orders}

We give an algorithm that decides in polynomial time whether a finite
ternary structure $S=(V,B)$ is $Bet(P)$ for some poset $P$, and produces one
if possible.\ 

\textbf{Lemma 13 }: Let $P=(V,\leq )$ is a finite and B-minimal partial
order.\ For $x,y\in V$, we have:

\begin{quote}
$x<y$ if and only if

either $B_{P}(x,y,z)$ holds for some $z\in Max(P),$ or

$y\in Max(P)$ and $B_{P}(w,x,y)\vee B_{P}(x,w,y)$ holds for some $w\in V$.
\end{quote}

\textbf{Proof }: The "if" direction is clear, since $Max(P)$ is not empty.

For the converse, assume that $x<y.\ $The elements $x$ and $y$ belong to a
chain of size at least 3 with maximal element\ $z\in Max(P)$.\ If $y\neq z$
we have $B_{P}(x,y,z),$ otherwise $B_{P}(w,x,y)\vee B_{P}(x,w,y)$ for some $%
w\in V$ (the last case is for the case where $x$ is minimal). $\square $

In a ternary structure $S=(V,B)$ that satisfies Properties B1, B2 and B3, we
say that an element $x\in V$ is \emph{extremal} if $B(y,x,z)$ does not hold
for any $y,z.$ We let $Ext(S)$ be the graph whose vertex set is the set of
extremal elements denoted by $V_{Ext}$, and $u-v$ is an edge\ if and only if 
$B(u,w,v)$ holds for some $w$ (necessarly not in $V_{Ext}$). It follows that 
$Ext(S)$ is a subgraph of $Gf(S)$.

\textbf{Lemma 14 }: Let $P=(V,\leq )$ be a finite and B-minimal partial
order without isolated elements.\ The graph $Ext(Bet(P))$ is bipartite, with
bipartition $(Max(P),Min(P))$.\ It is connected if $P$ is.

\textbf{Proof:} Each element of $Max(P)\cup Min(P)$\ is extremal in $Bet(P)$%
.\ If $w\notin Max(P)\cup Min(P)$\ then $u<w<v$ for some $u,v$ and so $%
B_{P}(u,w,v)$ holds and $w$\ is not extremal in $Bet(P)$.

The set $Max(P)\cap Min(P)$\ is empty.\ If $u-v$ is an edge of $Ext(Bet(P))$%
, then $B_{P}(u,w,v)$ holds hence $u<w<v$ or $v<w<u$ and $u$ and $v$ cannot
be both in $Max(P)$\ or in $Min(P)$.\ Hence, $Ext(Bet(P))$ is bipartite with
bipartition $(Max(P),Min(P))$.\ 

Assume now that $P$ is connected.\ Consider a sequence of maximal chains $%
C_{1},C_{2},...,C_{n}$ such that, for each $i,$ $C_{i}$ and $C_{i+1}$\ merge
or join, as in the proof of Theorem 7.\ We have in $Ext(Bet(P))$ a path

\begin{quote}
$Min(C_{1})-Max(C_{1})=Max(C_{2})-Min(C_{2})=Min(C_{3})-....-Min(C_{n})$
\end{quote}

or vice-versa by exchanging $Min$ and $Max$.\ It follows that $Ext(Bet(P))$
is connected.$\square $

This lemma proves in particular that the odd $B$-cycles (cf.\ Definition
1(c)) are not $Bet(P)$ for any partial order, because\ their extremal
elements are $a_{1},...,a_{n}$ forming an odd cycle.

\textbf{Theorem 15}\ : There exists a polynomial time algorithm that decides
whether a finite ternary structure $S=(V,B)$ is the betweenness structure of
a partial order $P$, and produce one if possible.\ 

\textbf{Proof }: Let be given a finite ternary structure $S=(V,B)$, $%
n:=\left\vert V\right\vert $ and $m:=\left\vert B\right\vert =O(n^{3}).$

\emph{Step 1\ }: In order to eliminate trivial cases, the algorithm first
checks Properties B1,B2 and B3, and if they hold, it constructs the graph $%
Gf(S$) and determines its connected components.\ This step takes time $%
O(n+m) $.

Each connected component is then considered and still denoted by $S$.

\emph{Step 2}: The algorithm constructs the graph $Ext(S)$, checks if it is
bipartite and if so determines its bipartition $(V_{1},V_{2})$ that is
unique because $Ext(S)$ is connected.\ If $Ext(S)$ is not bipartite, the
algorithm stops and returns a negative answer. This step takes time $O(m)$.

\emph{Step 3\ }: By taking $V_{1}$ as intended set of maximal elements, the
algorithm define a binary relation $\leq $ as follows:

\begin{quote}
$x\leq y$ if and only if either $x=y$,

or $B(x,y,z)$ holds for some $z\in V_{1},$

or $y\in V_{1}$ and $B(w,x,y)\vee B(x,w,y)$ holds for some $w\in V-V_{1}.$
\end{quote}

This step take time $O(m)$.\ If $S=Bet(P)$ for some poset $P$, then $\leq $
is a B-minimal partial order such that $Bet(Q)=S$, where $Q:=(V,\leq )$

\emph{Step 4\ }: The algorithm verifies that $B=Bet(Q)$.\ This step takes
time $O(n^{3})$. If $B\neq Bet(Q)$, it can report a failure.$\square $

This algorithm can perhaps be made quicker by means of clever
data-structures.

\section{No finite first-order axiomatization of $\mathcal{B}_{PO}$.}

\textbf{Proposition 16}\ : The class $\mathcal{B}_{PO}$ is not FO\
definable, \emph{i.e.}, is not characterized by a single first-order
sentence.

We recall that it is by the conjunction of an infinite set of universal
first-order sentences \cite{Lih}.

\textbf{Proof }: The proof is based on a reduction to a first-order
definability result for languages.\ 

A word $w$ of length $p>0$ over the alphabet $\{a,b\}$ can be represented by
the relational structure $S(w):=([p],\leq ,A)$ whose domain $[p]$\ is the
set of \emph{positions} of letters, $\leq $ is the standard order and $%
A\subseteq \lbrack p]$ is the set of positions of letter $a$.\ Hence $%
B:=[p]-A$ is the set of positions of $b$. We let $suc(x,y)$ express that $y$
is the position following $x$; this relation is FO definable from $\leq $.

The language $K:=\{(ab)^{m}\mid m\geq 2\}$ is FO definable, which means that
there exists an FO sentence $\kappa $ such that, for every nonempty word $w$
over $\{a,b\},$ we have $w\in K$ if and only if\ $S(w)\models \kappa $.

We define an FO formula $\beta (x,y,z)$ relative to $S(w)$ that expresses
the following property of a triple $(x,y,z)$ of positions of a word $w$ in $%
K $ :

\begin{quote}
$x\in A\wedge suc(x,y)\wedge suc(y,z)$ or $z\in A\wedge suc(z,y)\wedge
suc(y,x)$ or

$x\in A\wedge suc(y,z)$, $x$ is the first position of $w$ and $z$ is the
last one or

$z\in A\wedge suc(y,x)$, $z$ is the first position of $w$ and $x$ is the
last one.
\end{quote}

These triples and the opposite ones (to satisfy Property B2) form a set $%
B_{w}\subseteq \lbrack 2m]^{3}$\ The structure $T(w):=([2m],B_{w})$ is thus
a $B$-cycle as in Definition 1(c).\ Its vertives $a_{1},...,a_{m}$ are the
odd positions, those of letter $a$, its vertives $b_{1},...,b_{m}$ are the
even positions, those of letter $b$.

It follows from Definition 1(c) and Lemma 14 that $T(w)\in $ $\mathcal{B}%
_{PO}$ if and only if $m$ is even.

Assume now that $\mathcal{B}_{PO}$ is axiomatized by a first-order sentence $%
\theta $, possibly not universal. This sentence can be translated into an
FO\ sentence $\varphi $ such that, for every word $w$ over $\{a,b\},$ we
have:

\begin{quote}
$S(w)\models \kappa \wedge \varphi $ if and only if\ $w\in K$ and $%
T(w)\models \theta $

if and only if $T(w)\in \mathcal{B}_{PO}$ if and only if $w=(ab)^{2n}$ for
some $n$.
\end{quote}

Then the language $L:=\{(ab)^{2n}\mid n\geq 1\}$ would be first-order
definable, which is not the case by a classical result due to McNaughton,
Papert and Sch\"{u}tzenberger, see \cite{Tho}.

We get a contradiction hence, the class $\mathcal{B}_{PO}$ is not
axiomatizable by a single first-order sentence.$\square $

\textbf{Acknowledgement}: I thank Maurice Pouzet for his comments from which
I obtained the easy proof of Proposition 16.

\end{document}